\documentclass[aps,
pra,
reprint,
amsmath, amssymb,
superscriptaddress]{revtex4-2}

\usepackage{lmodern}
\usepackage[utf8]{inputenc}

\usepackage{bm}
\usepackage[retainorgcmds]{IEEEtrantools}
\usepackage{graphicx}
\usepackage{mathrsfs}
\usepackage{amsmath}
\usepackage{amssymb}
\usepackage{color}
\usepackage{amsfonts}
\usepackage{times,txfonts}
\usepackage{nicefrac}
\usepackage[colorlinks=true,linkcolor=blue,urlcolor=blue,citecolor=blue,pdfusetitle]{hyperref}
\usepackage{physics}
\usepackage{soul}
\usepackage{dsfont}
\usepackage{geometry}
\usepackage[dvipsnames]{xcolor}
\definecolor{mypink}{rgb}{0.858, 0.188, 0.478}
\definecolor{bondiblue}{rgb}{0.0, 0.58, 0.71}
\definecolor{bleudefrance}{rgb}{0.19, 0.55, 0.91}
\usepackage[authormarkup=none]{changes}
\definechangesauthor[name={Adalberto}, color=red]{g}

\begin{document}

\title{Entropy production from maximum entropy principle: a unifying approach}
\date{\today}
\author{Adalberto D. Varizi}
    \affiliation{Departamento de Ci\^encias Exatas e Tecnol\'ogicas, Universidade Estadual de Santa Cruz, 45662-900, Ilh\'eus, Bahia, Brazil}
\author{Pedro S. Correia}
    \affiliation{Departamento de Ci\^encias Exatas e Tecnol\'ogicas, Universidade Estadual de Santa Cruz, 45662-900, Ilh\'eus, Bahia, Brazil}

\begin{abstract}

Entropy production is the crucial quantity characterizing irreversible phenomena and the second law of thermodynamics.
Yet, a ubiquitous definition eludes consensus.
Given that entropy production arises from incomplete access to information, in this work we use Jaynes' maximum entropy principle to establish a framework that brings together prominent and apparently conflicting definitions. More generally our definition of entropy production addresses any tomographically incomplete quantum measurement and/or the action of a quantum channel on a system.

\end{abstract}

\maketitle{}


\section{\label{sec:intro}Introduction}
 Irreversible processes are omnipresent in nature. Their quantitative specification is provided in terms of an entropy production~\cite{Kondepudi2014,Landi2021Rev}.
This characterization allows the formalization of the second law of thermodynamics and the examination of a plethora of nonequilibrium phenomena like fluctuation theorems~\cite{Jarzynski1997,Esposito2009,Campisi2011}, thermodynamic uncertainty relations~\cite{Barato2015,Horowitz2020,Salazar2023}, the erasure of information~\cite{Landauer1961,Bennett1973,Plenio2001,Goold2015a,Timpanaro2020}, the thermodynamic role of coherences~\cite{Lostaglio2015,Santos2019,Francica2019,Scandi2019,Miller2019,Varizi2021} and the operation of thermal machines~\cite{Izumida2014,Brandner2015,Wang2016,Strasberg2021a,Elouard2023}.
Furthermore, entropy production restricts state transformations and is an important monotone in thermodynamic resource theories~\cite{Brandao2015,Lostaglio2015,Cwikli2015}.
It also can be used to characterize phase transitions~\cite{Dorner2012,Mascarenhas2014,Varizi2020,Goes2020,Varizi2022} and the effect of measurements on quantum systems~\cite{Elouard2017a,Elouard2017}.

Despite its significance and applicability, a consensual definition of entropy production is still lacking.
Indeed, the several definitions in the literature depend, for instance, on whether the system is open or closed and on whether one has access or not to individual trajectories~\cite{Jarzynski1997,Esposito2009,Campisi2011,Braasch2023}.
It can start from a thermodynamic entropy function~\cite{Polkovnikov2011b,Strasberg2021}; from the definition of an entropy flux combined with an entropy change~\cite{Breuer2003,Breuer2007,Schnakenberg1976}; or from the identification of a nonnegative contribution to the latter~\cite{Esposito2010a,Ptaszynski2023,Elouard2023}.

In this work we make a step in the direction of generalization.
Entropy is produced as a consequence of one's inability to retrieve information.
Therefore, it emerges whenever an observer does not have access to a tomographically complete set of observables or cannot perform a measurement on the system state eigenbasis.
Crucially this comprises the usual system-environment split.

Explicitly, founded on the Maximum Entropy Principle (MEP)~\cite{Jaynes1957}, an observer measuring a limited set of observables assigns to the system an unbiased state $\varrho_\text{max-S}$ solely based on the available information from these measurements.
Generically, this state will differ from the system state $\rho$ \emph{as determined by an observer with tomographically complete access}.
We then define the entropy production as the relative entropy between $\rho$ and $\varrho_\text{max-S}$.

From this framework we recover several definitions of entropy production in quantum thermodynamics.
For instance, when considering an observer performing local measurements, we obtain the definition in~\cite{Esposito2010a,Landi2021Rev}.
If we regard an observer performing fine- or coarse-grained energy measurements, we recover, respectively, the diagonal~\cite{Polkovnikov2011b} and observational~\cite{Safranek2019,Strasberg2021,Buscemi2023} entropy productions as particular cases.

Fundamentally our procedure applies to any tomographically incomplete measurement.
Since $\varrho_\text{max-S}$ will depend on what observables are or are not being measured, our formula allows us to understand how specific observables and control levels affect the entropy production.

Besides, since quantum channels can be seen as resulting from nonselective measurements~\cite{Breuer2007,Wilde2017,Nielsen2010a}, our reasoning also defines the entropy production associated with their action.
As an example, we show our entropy production for a system subjected to complete dephasing matches the relative entropy of coherence~\cite{Baumgratz2014}.

This article is organized as follows. In Sec.~\ref{sec:MEP}, we present our main result: a definition of entropy production derived from MEP.
Section~\ref{sec:examp} discusses key particular scenarios that demonstrate our approach and enable us to derive prominent definitions of thermodynamic entropy productions in Sec.~\ref{sec:Thermo}.
Section~\ref{sec:SecLaw} provides a brief overview of the role of entropy production in the second law of thermodynamics.
In Sec.~\ref{sec:ManyOneMaps}, we examine the production of entropy in one-to-one and many-to-one quantum channels.
Finally, in Secs. \ref{sec:Limits} and \ref{sec:conclusion}, we discuss the limitations of our approach, its connection with the second law of thermodynamics, and offer final remarks on the generality and further applications of our results.


\section{\label{sec:MEP} Maximum Entropy State and Entropy Production}
Let $\mathcal{L}(\mathcal{H}_n)$ denote the set of linear operators acting on a Hilbert space $\mathcal{H}_n$ of dimension $n$ and $\rho\in\mathcal{L}(\mathcal{H}_D)$ denote a system state according to a \emph{complete tomography}.
Unless $D$ is very small, a realistic observer has control over a limited number of degrees of freedom of this system.
Or yet, is able to measure a limited set of observables (Hermitian operators) $\{X_j\}$.
Consider an observer whose limited knowledge of this system is expressed by a set of measured numbers $\{x_j\}$ corresponding to the expected values of $\{X_j\}$: $x_j=\tr{X_j\rho}$. 

Furthermore, let us assume this system goes through a process described by a known completely positive and trace preserving (CPTP) map~\cite{Nielsen2010a,Wilde2017} $\Lambda:\mathcal{L}(\mathcal{H}_D)\rightarrow \mathcal{L}(\mathcal{H}_d)$, $D\geqslant d$, resulting in the output state $\Lambda(\rho)\in\mathcal{L}(\mathcal{H}_d)$.

Generically, after the process represented by $\Lambda$, the observer acquires information about the output state $\Lambda(\rho)$.
Let $\{O_i\}$ be a set of linear operators allowing the obtaining of such information through the experimental determination of the expected values $o_i=\tr{O_i\Lambda(\rho)}$.
Not knowing beforehand the system input state, but in possession of knowledge expressed by the constraints $\{o_i;x_j\}$, the observer may assign to the input system a compatible state $\varrho$ following some criteria~\cite{Zubarev1996,Vallejos2022,Jaynes1957,Buzek1998}.
The unique assignment consistent with all information available, while avoiding any bias, is the Maximum Entropy State (MES) upholding this set of constraints~\cite{Jaynes1957,Buzek1998,Zubarev1996}.

Formally, this goes as follows: The principle of maximum entropy dictates that we assign to the input system the state $\varrho$ with maximum von Neumann entropy $S(\varrho)=-\tr{\varrho\ln\varrho}$, subjected to the constraints $\{o_i;x_j\}$.
The solution of this problem involves finding the stationary point of the Lagrangian function $\mathcal{L}(\varrho;\{\lambda_i, \xi_i\}) = S(\varrho) - \sum_j \xi_j \left(\tr{X_j\varrho}-x_j\right) - \sum_i \lambda_i \left(\tr{O_i\Lambda(\varrho)}-o_i\right)$, where $\{\lambda_i\}$ and $\{\xi_j\}$ are Lagrange multipliers.

Let $\Lambda^*$ denote the adjoint of $\Lambda$, defined by $\tr{\Lambda^*(O_i)\rho}=\tr{O_i\Lambda(\rho)}$.
Then the state maximizing entropy and abiding to all constraints is given by~\cite{Buzek1998,Jaynes1957,Zubarev1996,Vallejos2022}
\begin{equation}\label{eq:maxS-S}
    \varrho_{\text{max-S}}^{\{o_i;x_j\}} = \frac{1}{Z}  \exp( - \sum_j\xi_jX_j - \sum_i \lambda_i\Lambda^*\Big(O_i\Big) ),
\end{equation}
where $Z=\tr{\exp( - \sum_j\xi_jX_j -\sum_i \lambda_i\Lambda^*\Big(O_i\Big) )}$ normalizes $\varrho_{\text{max-S}}^{\{o_i;x_j\}}$. 
The relation between $\xi_j$ and the associated constraint $x_j$ reads $x_j = - \frac{\partial}{\partial \xi_j}\ln Z$ and must be such that $\varrho_\text{max-S}^{\{o_i; x_j\}}$ predicts the correct measured expected value $x_j=\tr{X_j\varrho_\text{max-S}^{\{o_i; x_j\}}}$.
Similarly $\lambda_i$, implicitly given by $o_i = - \frac{\partial}{\partial \lambda_i}\ln Z$, is such that $\tr{O_i\Lambda(\varrho_\text{max-S}^{\{o_i; x_j\}})} = o_i$.

Now, whenever the constraints $\{o_i;x_j\}$ are insufficient to tomographically characterize the system, there must exist an entropy production.
In this case, given the input state of the system $\rho$~\footnote{Again as determined by an observer with tomographically complete access to the system.}, we can compare it with the unbiased \emph{assignment} $\varrho_\text{max-S}^{\{o_i,x_j\}}$ to obtain the entropy production associated with the available knowledge $\{o_i;x_j\}$.
Precisely, we define this entropy production by
\begin{equation}\label{eq:cgentprod}
    \Sigma^{\{o_i,x_j\}} = S\Big(\rho\,||\,\varrho_\text{max-S}^{\{o_i,x_j\}}\Big),
\end{equation}
where $S(\rho||\sigma)=\tr{\rho(\ln\rho-\ln\sigma)}\geqslant0$ is the quantum relative entropy.

Equation~\eqref{eq:cgentprod} constitutes our main result.
By construction, we see this definition is conveniently adjustable to distinct scenarios, with distinct accessible data.


\section{\label{sec:examp} Notable Examples}

In this section we explore key examples of Eqs.~\eqref{eq:maxS-S} and~\eqref{eq:cgentprod}, which we use in Sec.~\ref{sec:Thermo} to derive some prominent definitions of thermodynamic entropy productions from our main result.

Suppose an experimenter performs a measurement described by the \emph{complete} and \emph{orthogonal} set of rank-$1$ projectors $\{|a\rangle\langle a|\}$ associated with an observable $A$.

Let $\rho$ denote the state of the system as characterized by a full tomography.
If $\rho$ is not diagonal in the basis $\{|a\rangle\}$, this single measurement is not sufficient to completely determine the system state.
What this experimenter determines, nevertheless, is the set of populations $p_a=\tr{|a\rangle\langle a| \rho}$.
From this available knowledge, one might try to infer the state of the system.
The only unbiased inference is the MES consistent with the constraints $\{p_a\}$.
In connection with Eq.~\eqref{eq:maxS-S} we regard $\{|a\rangle\langle a|\}$ and $\{p_a\}$ as the sets of linear operators $\{X_j\}$ and constraints $\{x_j\}$~\footnote{Here the system does not go through a process/channel $\Lambda$, hence we discount this part in the maximum entropy state. In other words, since in this case the only constraints are the set $x_j=\tr{X_j\varrho}$, the MES have the form $\varrho_\text{max-S}^{\{x_j\}}=\exp(-\sum_j\xi_j X_j)/\tr{\exp(-\sum_j\xi_j X_j)}$.}.
This leads to the following MES~\eqref{eq:maxSFineMeasure}
\begin{equation}\label{eq:dephstate}
    \varrho_\text{max-S}^{\{p_a\}}=\sum_a p_a|a\rangle\langle a|.
\end{equation}

Since the full characterization of the system is given by $\rho$, there is an entropy production associated with the incomplete knowledge of the observer measuring $A$ that reads
\begin{equation}\label{eq:relcoh}
    \Sigma^{\{p_a\}} = S(\rho||\varrho_\text{max-S}^{\{p_a\}})
    \\[0.2cm]
    = S_{A}(\rho)-S(\rho),
\end{equation}
where $S_{A}(\rho) \equiv -\sum_a p_a\ln p_a$ is the so-called diagonal entropy of $\rho$ in the basis $\{|a\rangle\}$~\cite{Polkovnikov2011b}.

In Appendix~\ref{app:I} we show we can use Eqs.~\eqref{eq:maxS-S} and~\eqref{eq:cgentprod} to compute the entropy production associated with the completely dephasing map: $\mathds{D}_{A}(\rho)=\sum_a |a\rangle\langle a|\rho|a\rangle\langle a|$.
Not coincidentally, this entropy production equals~\eqref{eq:relcoh}.
The reason is because $\mathds{D}_{A}$ can be regarded as originating from the above measurement when the outcome is nonselected~\cite{Breuer2007,Wilde2017,Nielsen2010a}.
In this case, Eq.~\eqref{eq:relcoh} equals the \emph{relative entropy of coherence}~\cite{Baumgratz2014,Streltsov2016a} and quantifies the entropy production due to the loss of $\{|a\rangle\}$-basis coherences in $\rho$ enforced by the completely dephasing process.

Consider now an observer performing a \emph{coarse-grained measurement}~\cite{Safranek2019,Strasberg2021} with a family of projectors $\Pi_i=\sum_\mu|a_{i\mu}\rangle\langle a_{i\mu}|$, with rank $V_i=\tr{\Pi_i}\geqslant1$, which are orthogonal, $\Pi_i\Pi_j=\delta_{ij}\Pi_i$, and complete, $\sum_i \Pi_i=\mathds{1}$.
As motivation, we imagine the observable $A=\sum_i\sum_\mu a_{i\mu}|a_{i\mu}\rangle\langle a_{i\mu}|$ as having blocks of size $V_i$ of nearly degenerate eigenvalues $\{a_{i\mu}\}$, such that this observer cannot experimentally resolve between eigenvalues in the same block.
This is necessarily the case in a macroscopic system, where, for instance, the separation between energy levels is exponentially small in the system number of particles~\cite{Landau-Lif1980,Reimann2010}.

Denoting again by $\rho$ the full tomographic characterization of the system, this measurement allows the acquisition of the probabilities $p_i=\tr{\Pi_i\rho}$ of finding the system in subspace $\Pi_i$.
Preceding as in the first example -- with $\{p_i\}$ and $\{\Pi_i\}$ in place of $\{x_j\}$ and $\{X_j\}$ -- an inference of the system state based on the MEP leads to~\eqref{eq:maxSCoarseMeasureClose}
\begin{equation}\label{eq:obsstat}
    \varrho_\text{max-S}^{\{p_i\}} = \sum_i p_i \frac{\Pi_i}{V_i}.
\end{equation}

Hence, the incomplete determination of the system state in this case results in an entropy production given by~\eqref{APPeq:obsent1}
\begin{equation}\label{eq:obsent}
    \Sigma^{\{p_i\}}=S(\rho||\varrho_\text{max-S}^{\{p_i\}})= S_\text{obs}^{\{\Pi_i\}}(\rho) - S(\rho),
\end{equation}
where $S_\text{obs}^{\{\Pi_i\}}(\rho)= - \sum_i p_i\ln (p_i/V_i)$ is the so-called \emph{observational entropy} of $\rho$~\cite{Safranek2019,Strasberg2021,Buscemi2023}.
We might think of this entropy production as resulting from lack of knowledge of the observer about the precise population of each state $|a_{i\mu}\rangle$ and about the coherences between these states.
In Appendix~\ref{app:II} we also consider a quantum channel leading to the same entropy production.

Moving forward, we consider henceforth an open system $S$ interacting with an environment $E$.
The reason for splitting the global system into parts $S$ and $E$ is the assumption that an observer has access only to local operations in the former; hence, only acquires information about $S$.
Equivalently we might think of the global state $\rho_{SE}$ subjected to a process represented by the partial trace over $E$ channel, resulting in the reduced state $\rho_S=\tr_E\{\rho_{SE}\}$.

Let us assume $\rho_S$ to be fully known by the observer, meaning she/he has access to a tomographic complete set of observables $\{O_i^S\}$ with known expected values $o_i^S=\tr{O_i^S\rho_S}$.
Note that $\tr_E^*(O_i^S)=O_i^S\otimes\mathds{1}_E$, where $\mathds{1}_E$ is the identity over the environment.
Applying~\eqref{eq:maxS-S}, the MES associated with the action of $\tr_E$ under the constraints~$\{o_i^S\}$ is --- see Appendix~\ref{app:III} and~\cite{Vallejos2022},
\begin{equation}
    \varrho_\text{max-S}^{\{o_i^S\}} = \rho_S\otimes\frac{\mathds{1}_E}{d_E},
\end{equation}
where $d_E=\tr{\mathds{1}_E}$ is the dimension of the environment Hilbert space, and we used that $\rho_S$ is fixed by the constraints $\{o_i^S\}$.

Hence, the discarding of $E$ -- represented by the partial trace -- when the reduced state of $S$ is known to be $\rho_S$ produces entropy by an amount
\begin{equation}\label{eq:ent-E1}
    \Sigma^{\{o_i^S\}} = S(\rho_{SE}||\rho_S\otimes\mathds{1}_E/d_E).
\end{equation}

Essentially these examples and those in the appendices illustrate the versatility of Eq.~\eqref{eq:cgentprod} and how it adapts to different scenarios.
The first two examples show how different control levels affect the entropy production.
The effects of access to different observables also become clear below when we consider thermodynamic processes and assume an observer with some knowledge of the environment.


\section{\label{sec:Thermo} Applications to Thermodynamics}
Building on the results of the previous section, we now demonstrate how our approach recovers the diagonal~\cite{Polkovnikov2011b} and observational~\cite{Safranek2019,Strasberg2021,Buscemi2023} definitions of thermodynamic entropy production as particular cases for when an observer performs fine- or coarse-grained energy measurements, respectively. Additionally, we show that, when considering an observer performing local measurements, we obtain the definition in~\cite{Esposito2010a,Landi2021Rev}.

Consider a system evolving unitarily driven by a time-dependent Hamiltonian $H(t)=\sum_i \epsilon_i^t|\epsilon_i^t\rangle\langle\epsilon_i^t|$.
Let $U_t=\mathcal{T}\exp{i\int_0^t H(t')\mathrm{d}t'}$ denote the unitary time-evolution operator up to time $t$ --- here $\mathcal{T}$ represents the time-ordering operator.
We assume the system is initially in a state $\rho_0=\sum_i p_i^0|\epsilon_i^0\rangle\langle\epsilon_i^0|$, diagonal in the initial energy basis.
This means a measurement in this basis completely determines $\rho_0$~\footnote{Since $\rho_0$ is diagonal in the energy basis, the system state is completely characterized by the populations $p_i^0$, which can be obtained solely from an energy measurement}.
It further means $S_{H_0}(\rho_0)=S(\rho_0)$ --- i.e., the diagonal entropy of $\rho_0$ in the $H_0$ basis equals its von Neumann entropy.

Consider that after an evolution up to time $t$, the observer measures the system in the current energy basis $\{|\epsilon_i^t\rangle\}$.
In general, $\rho_t=U_t\rho_0U_t^\dagger$ will not be diagonal in this basis and, therefore, this measurement cannot tomographically identify the system state.
As seen in the previous section -- Eq.~\eqref{eq:relcoh} -- this leads to an entropy production given by
\begin{equation}\label{eq:d-entprod}
\begin{aligned}
    \Sigma^\text{d} &= S_{H(t)}(\rho_t)-S(\rho_t) 
    \\[0.2cm]
    &= S_{H(t)}(\rho_t)-S_{H_0}(\rho_0),
\end{aligned}
\end{equation}
where we used $S(\rho_t)=S(\rho_0)=S_{H_0}(\rho_0)$.

In~\cite{Polkovnikov2011b} Polkovnikov proposes the use of diagonal entropy as the thermodynamic entropy.
He then showed that for a closed system initially diagonal in the energy basis we would have $S_{H(t)}(\rho_t)\geqslant S_{H_0}(\rho_0)$, meaning the thermodynamic entropy of the system had increased ---  agreeing with the second law.
In this case, the entropy produced would be precisely~\eqref{eq:d-entprod}, which we can see here as a particular case of our definition in~\eqref{eq:cgentprod}.

Next, we continue considering a closed system evolving unitarily.
But now, instead of a fully-resolved, we consider a coarse-grained energy measurement with the complete and orthogonal set of projectors $\{\Pi_i^t=\sum_\mu |\epsilon_{i\mu}^t\rangle\langle\epsilon_{i\mu}^t|\}$, where we write the system Hamiltonian as $H(t)=\sum_{i\mu}\epsilon_{i\mu}^t|\epsilon_{i\mu}^t\rangle\langle\epsilon_{i\mu}^t|$.
We assume the system's initial state is $\rho_0 = \sum_i p_i^0\Pi_i^0/V_i^0$, so that a measurement with $\{\Pi_i^0\}$ is sufficient to completely determine $\rho_0$.
Consequently its observational and von Neumann entropies equal: $S_\text{obs}^{\{\Pi_i^0\}}(\rho_0)=S(\rho_0)$.

Again, let $\rho_t=U_t\rho_0U_t^\dagger$ denote system state evolved up to time $t$, when a measurement with the up-to-the-time projectors $\{\Pi_i^t\}$ is performed to probe the system.
As seen above --- Eq.~\eqref{eq:obsent}, this generically leads to an entropy production given by,
\begin{equation}\label{eq:obs-entprod}
\begin{aligned}
     \Sigma^\text{obs} &=S_\text{obs}^{\{\Pi_i^t\}}(\rho_t)-S_\text{obs}^{\{\Pi_i^0\}}(\rho_0),
\end{aligned}
\end{equation}
where we used $S(\rho_t)=S(\rho_0)=S_\text{obs}^{\{\Pi_i^0\}}(\rho_0)$.

Equation~\eqref{eq:obs-entprod} is the definition of thermodynamic entropy production for a unitarily evolving system in~\cite{Strasberg2021}.
In~\cite{Safranek2019,Strasberg2021} the authors propose the use of observational entropy as the thermodynamic entropy since it interpolates between the von Neumann and Boltzmann entropies.
We notice this definition of entropy production follows directly from Eq.~\eqref{eq:cgentprod} by considering the appropriate scenario.

Continuing the applications of Eq.~\eqref{eq:cgentprod}, we assume henceforth an open system $S$ interacting with an environment $E$.
Typically one considers the situation where the system, in a state $\rho_S^0$, is put to interact with the environment in a state $\rho_E^0$.
The global state then evolves unitarily to $\rho_{SE}=U_{SE}(\rho_S^0\otimes\rho_E^0)U_{SE}^\dagger$.
At the end of the interaction, the environment is disregarded and a set of local measurements is performed to determine the system's reduced state $\rho_S=\tr_E\{\rho_{SE}\}$.

Equation~\eqref{eq:ent-E1} computes the entropy produced in the process when nothing is known about the environment.
Usually, however, that is not the case.
Suppose, for instance, the environment initial energy $E_E^0$ is known.
In Appendix~\ref{app:III} we show we may use this as a further constraint leading to the MES $\varrho_\text{max-S}^{\{o_i^S;E_E^0\}} = \rho_S\otimes e^{-\beta_0 H_E}/Z_E$ and the entropy production $\Sigma^{\{o_i^S;E_E^0\}} = \Delta S_S + \beta_0\Delta E_E$.
Here $\Delta S_S=S(\rho_S)-S(\rho_S^0)$ is the change in von Neumann entropy of the system, $\Delta E_E=\tr{(\rho_E-\rho_E^0)H_E}$ is the change in energy of the environment and $\beta_0$ is defined by $E_E^0=-\partial_{\beta_0}\ln Z_E$, where $Z_E=\tr\{e^{-\beta_0H_E}\}$.
According to~\cite{Esposito2010a,Landi2021Rev}, this is exactly the entropy produced in a process in which the environment is initially in a thermal state $\rho_E^0 = e^{-\beta_0 H_E}/Z_E$ at inverse temperature $\beta_0$.

Ultimately, if the set of constraints $\Big\{o_j^{E_0}=\tr{O_j^E\rho_E^0}\Big\}$ completely specifies the environment initial state $\rho_E^0$ we can use them in Eq.~\eqref{eq:maxS-S} to obtain $\varrho_\text{max-S}^{\{o_i^S;o_j^{E_0}\}}=\rho_S\otimes\rho_E^0$~\eqref{eq:maxSQSTOpen}.
This leads to the general definition of entropy production in~\cite{Esposito2010a,Landi2021Rev}:
\begin{equation}\label{eq:entprod}
    \Sigma^\text{info} = S(\rho_{SE}||\rho_S\otimes\rho_E^0) = S(\rho_E||\rho_E^0) + \mathcal{I}_{SE}(t),
\end{equation}
where $\rho_E=\tr_S\{\rho_{SE}\}$ is the final environment state and $\mathcal{I}_{SE}(t)=S(\rho_{SE}||\rho_S\otimes\rho_E)$ is the quantum mutual information, quantifying the total amount of correlations in the global state $\rho_{SE}$.
The last equality in~\eqref{eq:entprod} shows this entropy production emerges from the loss of information about the final environment state and the correlations between $S$ and $E$ when the environment is discarded.
This discussion reveals how access to different observables, here particularly of $E$, modifies the entropy production.
In Appendix~\ref{app:III} we further show that when the observer knows the final (local) state of the environment the entropy production reduces to $I_{SE}(t)$.

Moving forward, we consider now an observer who can perform only coarse-grained energy measurements on a large environment.
Particularly, we denote by $\{|s_i^t\rangle\langle s_i^t|\}$ the eigenprojectors of the reduced system state at any time $t$; by $H_E=\sum_{j\mu} \epsilon_{j\mu}^E|\epsilon_{j\mu}^E\rangle\langle\epsilon_{j\mu}^E|$ the environment Hamiltonian and by $\{\Pi_j^E=\sum_\mu |\epsilon_{j\mu}^E\rangle\langle\epsilon_{j\mu}^E|\}$ the complete and orthogonal set of energy projectors used to probe $E$. For simplicity, we take $H_E$ to be time-independent.
We assume the observer can perform, at any time $t$, a joint measurement of system and environment described by the set $\{|s_i^t\rangle\langle s_i^t|\otimes\Pi_j^E\}$.
Moreover, we assume the initial joint global state to be of the form $\rho_{SE}^0=\sum_{ij} s_i^0|s_i^0\rangle\langle s_i^0|\otimes p_j^0\Pi_j^E/V_j^E$, with $V_j^E=\tr{\Pi_j^E}$, such that its observational entropy equals its von Neuman entropy: $S_\text{obs}^{\{|s_i^0\rangle\langle s_i^0|\otimes\Pi_j^E\}}(\rho_{SE}^0)=S(\rho_{SE}^0)$.

After a global unitary evolution leading to the final state $\rho_{SE}=U_{SE}\rho_{SE}^0U_{SE}^\dagger$ at time $t$, the observer performs the aforementioned joint coarse-grained measurement.
This allows the determination of the probabilities $p_{ij}=\tr{|s_i^t\rangle\langle s_i^t|\otimes\Pi_j^E\rho_{SE}}$ of finding the system in state $|s_i^t\rangle$ while the environment is in the energy subspace $\Pi_j^E$.
From them the MES associated with the available knowledge is~\eqref{APPeq:ObsStat2}
\begin{equation}
    \varrho_\text{max-S}^{\{p_{ij}\}} = \sum_{ij} p_{ij}|s_i^t\rangle\langle s_i^t| \otimes \Pi_j^E/V_j^E.
\end{equation}

The entropy production in this case becomes --- see Eq.~\eqref{eq:obsent}:
\begin{equation}\label{eq:obs-entprod2}
\begin{aligned}
    \Sigma^\text{obs} 
    &= S_\text{obs}^{\{|s_i^t\rangle\langle s_i^t|\otimes\Pi_j^E\}}(\rho_{SE}) - S_\text{obs}^{\{|s_i^0\rangle\langle s_i^0|\otimes\Pi_j^E\}}(\rho_{SE}^0),
\end{aligned}
\end{equation}
where $S_\text{obs}^{\{|s_i^t\rangle\langle s_i^t|\otimes\Pi_j^E\}}(\rho_{SE})=-\sum_{ij} p_{ij}\ln(p_{ij}/V_j^E)$ is the observational entropy of the final state and we used $S(\rho_{SE})=S(\rho_{SE}^0)$.
Equation~\eqref{eq:obs-entprod2} constitutes the definition of entropy production for an open system in~\cite{Strasberg2021} based on observational entropy.
In Appendix~\ref{app:IV} we further discuss the case of independent local measurements on $S$ and $E$.

Hence, several prominent definitions of thermodynamic entropy production are encapsulated in the MEP combined with Eq.~\eqref{eq:cgentprod}.
Our approach operationally unveils how these distinct definitions arise from different assumptions over what is being measured.


\section{\label{sec:SecLaw}Entropy Production and the Second Law of Thermodynamics}
Broadly there are currently two views on the meaning of the second law of thermodynamics.
On one hand, many consider that given a system with its specific properties and dynamics, any identification of a strictly nonnegative contribution to the change in its von Neumann entropy constitutes a statement of the second law.
This is the position adopted implicitly or explicitly for instance in~\cite{Breuer2003,Landi2021Rev,Esposito2010a,Ptaszynski2023,Elouard2023}.
In this case, Eq.~\eqref{eq:cgentprod} can be seen as a generalized statement of the second law.

Specifically, regarding a system-environment in an initial state $\rho_S^0\otimes\rho_E^0$ evolving unitarily, we may write $\Delta S_S = \Sigma + \Phi$.
Here, $\Delta S_\alpha$ is the change in von Neumann entropy of $\alpha$ and $\Sigma = S(\rho_{SE}||\varrho_\text{max-S})$ is the entropy production in the process.
$\Phi= - \Delta S_E - \tr{\rho_{SE}(\ln\rho_S\otimes\rho_E-\ln\varrho_\text{max-S})}$ is the entropy flux into/out of $S$, having no definite sign.
Hence, $\Sigma$ in Eq.~\eqref{eq:cgentprod} gives the strictly nonnegative contribution to the change in entropy of the system.

Contrastingly, some assume any statement of the second law necessarily starts from the definition of a strictly nondecreasing thermodynamic entropy function, applicable also for closed systems.
This position is endorsed for instance by~\cite{Polkovnikov2011b,Safranek2019,Strasberg2021,Strasberg2021a}.
In this view, the von Neumann entropy cannot, in general, be used as the thermodynamic entropy because of its invariance in closed systems.
One then must adopt some other function such as diagonal or observational entropies.
In this case, Eq.~\eqref{eq:cgentprod} serves to quantify the increase in this function.

Interestingly, the increase in diagonal and observational entropies in Eqs.~\eqref{eq:d-entprod} and~\eqref{eq:obs-entprod} are equivalent to the increase in \emph{von Neumann} entropy of the respective maximum entropy states
This suggests we can always use the von Neumann entropy as the thermodynamic entropy, as long as we apply it to the unbiased state encoding \emph{only} and \emph{all available} information about the system: $\varrho_\text{max-S}$.
This advocates further the interpretation of $\varrho_\text{max-S}$ as a macrostate representation of the system, specifying solely the expected values of the measurable physical quantities in a given setup.


\section{\label{sec:ManyOneMaps}Many-to-One and One-to-One Channels}
We already mentioned, and show in~\ref{app:I} and~\ref{app:III}, how Eqs.~\eqref{eq:relcoh} and~\eqref{eq:ent-E1} may be seen as the entropy production resulting from the full dephasing and partial trace operations, respectively.
Indeed, equations~\eqref{eq:maxS-S} and~\eqref{eq:cgentprod} are directly applicable to \emph{many-to-one} channels: those $\Lambda$ for which \emph{many} input $\rho$ lead to the same output $\Lambda(\rho)$ --- in~\ref{app:II} we give a further example of such a channel in connection with Eq;~\eqref{eq:obsent}.
For such channels, knowledge of the output state generically does not determine the input.
In other words, $\varrho_\text{max-S}\neq\rho$, leading to a nonzero entropy production characterizing the irreversibility of the process $\Lambda$.

Still, many important quantum channels are of the type one-to-one --- these include noisy channels like the bit-flip, phase-flip, depolarizing and amplitude damping~\cite{Wilde2017,Nielsen2010a}.
For these channels, direct application of equations~\eqref{eq:maxS-S} and~\eqref{eq:cgentprod} leads to a zero entropy production.
This is because $\varrho_\text{max-S}$ must be such that $\Lambda(\varrho_\text{max-S})=\Lambda(\rho)$.
Hence, a one-to-one relation between the input and output of $\Lambda$ forces $\varrho_\text{max-S}=\rho$, culminating in Eq.~\eqref{eq:cgentprod} vanishing.
Nevertheless, in Appendix~\ref{app:V} we show we can still use Eq.~\eqref{eq:cgentprod} to compute the change in von Neumann entropy of the system after passing through a one-to-one channel.


\section{\label{sec:Limits}Limitations}
The MES in Eq.~\eqref{eq:maxS-S} is valid for constraints given by the expected values of linear operators.
Some definitions of entropy production do not fit this paradigm~\cite{Elouard2023,Ptaszynski2023}.
For example, in~\cite{Elouard2023} the authors consider the system-environment state $\rho_S^0\otimes\rho_E^0$ evolving unitarily.
They then assume knowledge of the initial and final von Neumann entropies of the environment and use them to define nonequilibrium initial and final temperatures for $E$.
Particularly, they define the nonequilibrium inverse temperature of a system with Hamiltonian $H$ in a state $\rho$ as the number $\beta$ such that $S(\rho)=S(\rho^\beta)$, where $\rho^\beta=e^{-\beta H}/\tr{e^{-\beta H}}$.
This allows the introduction of a so-called thermal energy $E^\text{th}(t)=\tr{H\rho^{\beta(t)}}$, which they use to define an entropy production: $\Sigma=\Delta S_S +\beta_0\Delta E_E^\text{th}(t)=I_{SE}(t)+S(\rho_E^{\beta(t)}||\rho_E^{\beta_0})$, where $\Delta S_S$ is the change in von Neumann entropy of the system and $\Delta E_E^\text{th}(t)$ is the change in thermal energy of the environment.

Since the von Neumann entropy of a state cannot be written as the expected value of a linear operator, such a definition of entropy production cannot be derived from our framework.
We note, however, that computing the entropy of the environment final state generally requires full tomography of this state.
As aforementioned, is possible to show using Eq.~\eqref{eq:cgentprod}, though, that full knowledge of the final environment state should lead to a smaller entropy production, given solely by $I_{SE}(t)$.  


\section{\label{sec:conclusion}Conclusion}
Understanding the emergence of thermodynamics, especially the second law, from the microscopic description of systems is a fundamental goal in quantum thermodynamics.
Many definitions of entropy production --- the quantity identifying irreversibility --- have been proposed already.
Here we presented a resourceful framework grounded on the maximum entropy principle that reconciles several prominent proposals.

Beyond that, our approach operationally connects entropy production with the accessible properties of a system.
This is achieved through the introduction of a macrostate~\eqref{eq:maxS-S} of Gibbs-like form, suggesting deep relations with thermodynamics.
Particularly, we expect the identification of Lagrange multipliers in~\eqref{eq:maxS-S} with standard thermodynamic variables.
Indeed, a nonequilibrium temperature based on equality of a system internal energy with that of a fictitious Gibbs state is used in~\cite{Strasberg2021}.
Such definition emerges naturally in our framework.

Exploring the link between quantum measurements and maps~\cite{Breuer2007,Wilde2017,Nielsen2010a}, our definition of entropy production extends to general quantum channels.
In this direction, we hope to explore the entropy production in systems whose degrees of freedom of interest cannot be split in the usual system-environment form~\cite{Duarte2017,Correia2019,Correia2021,Vallejos2022,Correia2023}.


\section*{Acknowledgement}
ADV acknowledges financial support from the Coordena\c c\~ao de Aperfei\c coamento de Pessoal de N\'ivel  Superior - Brasil (CAPES). PSC acknowledges financial support from the National Council for Scientific and Technological Development, CNPq Brazil (projects: Universal Grant No. 406499/2021-7, and 409611/2022-0)

\appendix

\section{\label{app:I}Entropy Production in Fine-Grained Measurement and Dephasing Channel}

As in the main text, we consider here a measurement described by the projectors set $\{|a\rangle\langle a|\}$ such that $\tr{|a\rangle\langle a|}=1$, $\langle a|a'\rangle=\delta_{a,a'}$ and $\sum_a |a\rangle\langle a|=\mathds{1}$.
Given a system preparation $\rho$, this measurement allows the determination of the probabilities $p_a=\tr{|a\rangle\langle a|\rho}=\langle a|\rho|a\rangle$.

Referring to Eqs.~\eqref{eq:maxS-S} and~\eqref{eq:cgentprod}, here the set $\{p_a\}$, associated with the linear operators $\{|a\rangle\langle a|\}$ corresponds to the constraints $\{x_j\}$ representing knowledge of the observer about the system.
Using the orthogonality of the projectors $\{|a\rangle\langle a|\}$, we have that the state of maximum entropy consistent with this knowledge is given by
\begin{equation}
    \varrho_\text{max-S}^{\{p_a\}}=\frac{1}{Z}\exp(-\sum_a \xi_a |a\rangle\langle a|) = \sum_a \frac{e^{-\xi_a}}{Z} |a\rangle\langle a|,
\end{equation}
where
\begin{equation}
    Z=\tr{\exp(-\sum_a \xi_a |a\rangle\langle a|)} = \sum_a e^{-\xi_a},
\end{equation}
and the Lagrange multipliers $\{\xi_a\}$ must be such that
\begin{equation}
    p_a \equiv -\frac{\partial}{\partial\xi_a}\ln Z = -\frac{\partial}{\partial\xi_a}\ln \sum_a e^{-\xi_a} = \frac{e^{-\xi_a}}{Z}.
\end{equation}
Hence,
\begin{equation}\label{eq:maxSFineMeasure}
    \varrho_\text{max-S}^{\{p_a\}}=\sum_a p_a|a\rangle\langle a|.
\end{equation}

Accordingly, the accompanying entropy production is given by
\begin{equation}\label{APPeq:relcoh1}
\begin{aligned}
    \Sigma^{\{p_a\}} &= S\left(\rho||\varrho_\text{max-S}^{\{p_a\}}\right)=\tr{\rho\left(\ln \rho - \ln \varrho_\text{max-S}^{\{p_a\}}\right)} 
    \\[0.2cm]
    &= S_A(\rho) - S(\rho),
\end{aligned}
\end{equation}
where $S(\rho)=-\tr{\rho\ln\rho}$ is the von Neumann entropy of $\rho$ and
\begin{equation}
    S_A(\rho) = - \tr{\rho \ln \sum_a p_a |a\rangle\langle a|} = - \sum_a p_a \ln p_a,
\end{equation}
is the diagonal entropy~\cite{Polkovnikov2011b} of $\rho$ in the basis $\{|a\rangle\}$.

As stated in the main text, Eq.~\eqref{APPeq:relcoh1} is also equal to the entropy production associated with the action of the (many-to-one) completely dephasing map:
\begin{equation}
    \mathds{D}_A(\rho) = \sum_a |a\rangle\langle a|\rho|a\rangle\langle a| = \sum_a p_a |a\rangle\langle a|.
\end{equation}

Let us assume an experimenter with access to a tomographically complete set of observables $\{O^i\}$ used to determine the output state $\mathds{D}_A(\rho)$.
The expected values of these observables are given by $o_i=\tr{O^i\mathds{D}_A(\rho)}=\sum_a p_a \langle a|O^i|a\rangle = \sum_a p_a O_{aa}^i$.
Furthermore, we assume this experimenter to have no information about the input state $\rho$.
It is easy to show the trace dual of the channel $\mathds{D}_A$ is $\mathds{D}_A$ itself.
Then the maximum entropy state associated with these constraints is given by
\begin{equation}
    \begin{aligned}
    \varrho_\text{max-S}^{\{o_i\}} &= \frac{1}{Z}\exp{-\sum_i \lambda_i \mathds{D}_A(O^i)} 
    \\[0.2cm]
    &= \frac{1}{Z}\exp{-\sum_{i, a} \lambda_i O_{aa}^i|a\rangle\langle a|}
    \\[0.2cm]
    &= \sum_a \frac{e^{- \sum_i \lambda_i O_{aa}^i}}{Z} |a\rangle\langle a|,
    \end{aligned}
\end{equation}
where
\begin{equation}
    Z=  \tr{\exp{-\sum_i \lambda_i \mathds{D}_A(O^i)}} = \sum_a e^{-\sum_i \lambda_i O_{aa}^i}.
\end{equation}
The Lagrange multipliers $\{\lambda_i\}$ must be such that
\begin{equation}
        o_i = \sum_a p_a O_{aa}^i \equiv -\frac{\partial}{\partial\lambda_i}\ln Z = \sum_a \frac{e^{-\sum_i \lambda_i O_{aa}^i}}{Z} O_{aa}^i.
\end{equation}
Hence, $e^{-\sum_i \lambda_i O_{aa}^i}/Z=p_a$ and
\begin{equation}
    \varrho_\text{max-S}^{\{o_i\}} = \sum_a p_a |a\rangle\langle a| = \mathds{D}_A(\rho).
\end{equation}

Consequently the entropy produced when the system passes through the dephasing process is given by~\eqref{APPeq:relcoh1}, which is equivalent to the relative entropy of coherence~\cite{Baumgratz2014} of $\rho$ in the basis $\{|a\rangle\}$.


\section{\label{app:II}Entropy Production in Coarse-Grained Measurement}

Let us consider now a coarse-grained measurement~\cite{Safranek2019,Strasberg2021} described by the set of projectors $\{\Pi_i\}$ with ranks $V_i=\tr{\Pi_i}$, satisfying the orthogonality condition $\Pi_i\Pi_j=\delta_{ij}\Pi_j$ and the completeness relation $\sum_i \Pi_i=\mathds{1}$.

For a system prepared in a state $\rho$, an observer performing this measurement  will acquire knowledge about the system in the probabilities $p_i=\tr{\Pi_i\rho}$.
The maximum entropy state constrained by this knowledge reads
\begin{equation}
    \varrho_\text{max-S}^{\{p_i\}}=\frac{1}{Z}\exp{-\sum_i \xi_i \Pi_i} = \sum_i \frac{e^{-\xi_i}}{Z}\Pi_i,
\end{equation}
with
\begin{equation}
    Z = \tr{\exp{-\sum_i \xi_i \Pi_i}} = \sum_i V_i e^{-\xi_i},
\end{equation}
where the Lagrange multipliers $\{\xi_i\}$ must be such that
\begin{equation}
    p_i \equiv - \frac{\partial}{\partial\xi_i}\ln Z = V_i\frac{e^{-\xi_i}}{Z}.
\end{equation}
Therefore, $e^{-\xi_i}/Z=p_i/V_i$ and
\begin{equation}\label{eq:maxSCoarseMeasureClose}
    \varrho_\text{max-S}^{\{p_i\}} = \sum_i p_i  \frac{\Pi_i}{V_i}.
\end{equation}

Thus, the entropy production associated with the limited knowledge about the system encoded in the constraints $\{p_i\}$ reads
\begin{equation}\label{APPeq:obsent1}
 \begin{aligned}
    \Sigma^{\{p_i\}} &= S(\rho||\rho_\text{max-S}^{\{p_i\}}) = \tr{\rho(\ln\rho - \ln\varrho_\text{max-S}^{\{p_i\}})} 
    \\[0.2cm]
    &= S_\text{obs}^{\{\Pi_i\}}(\rho) - S(\rho),
 \end{aligned}
\end{equation}
where
\begin{equation}
    S_\text{obs}^{\{\Pi_i\}}(\rho) = -\tr{\rho \ln \sum_i p_i \frac{\Pi_i}{V_i}} = -\sum_i p_i \ln \frac{p_i}{V_i}
\end{equation}
is the observational entropy~\cite{Safranek2019,Strasberg2021} of $\rho$ associated with the coarse-graining $\{\Pi_i\}$. 

There is also a many-to-one channel that leads to the same entropy production in~\eqref{APPeq:obsent1}.
Let us denote by $\{|a_{i\mu}\rangle\langle a_{i\mu}|\}$ the set of rank-$1$ projectors such that $\Pi_i=\sum_{\mu} |a_{i\mu}\rangle\langle a_{i\mu}|$.
We specify the many-to-one channel $\Lambda_\text{obs}$ by the Krauss operators:
\begin{equation}
    K_{i\mu\nu} = \frac{1}{\sqrt{V_i}} |a_{i\mu}\rangle\langle a_{i\nu}|.
\end{equation}

The action of $\Lambda_\text{obs}$ on state $\rho$ leads to the output
\begin{equation}
    \Lambda_\text{obs}(\rho) = \sum_{i\mu\nu}K_{i\mu\nu}\rho K_{i\mu\nu}^\dagger = \sum_i p_i \frac{\Pi_i}{V_i}.
\end{equation}

Let $\{O^\alpha\}$ denote the tomographically complete set of observables an experimenter uses to determine the output state $\Lambda(\rho)$.
The expected values of these observables on the final state read $o_\alpha=\tr{O^\alpha\Lambda_\text{obs}(\rho)}=\sum_i (p_i/V_i)O^\alpha_i$, where $O^\alpha_i=\tr{O^\alpha\Pi_i}$.
We assume an observer with no knowledge of the input $\rho$.
The action of the trace-dual of $\Lambda_\text{obs}$ is given by $\Lambda_\text{obs}^{*}(\bullet)=\sum_{i\mu\nu}K_{i\mu\nu}^\dagger(\bullet)K_{i\mu\nu}$.
The maximum entropy state associated with the constraints $\{o_\alpha\}$ is given by
\begin{equation}
 \begin{aligned}
    \varrho_\text{max-S}^{\{o_\alpha\}} &= \frac{1}{Z}\exp{-\sum_\alpha \lambda_\alpha \Lambda_\text{obs}^{*}(O^\alpha)} 
    \\[0.2cm]
    &= \frac{1}{Z}\exp{-\sum_\alpha \lambda_\alpha \sum_i O^\alpha_i \Pi_i/V_i} 
    \\[0.2cm]
    &= \sum_i \frac{e^{-(1/V_i)\sum_\alpha \lambda_\alpha O^\alpha_i}}{Z}\Pi_i,
 \end{aligned}
\end{equation}
where
\begin{equation}
    Z = \tr{\exp{-\sum_\alpha \lambda_\alpha \Lambda_\text{obs}^{*}(O^\alpha)}} = \sum_i V_i e^{-(1/V_i) \sum_\alpha \lambda_\alpha O^\alpha_i}.
\end{equation}
The Lagrange multipliers $\{\lambda_\alpha\}$ must be such that
\begin{equation}
    o_\alpha = \sum_i \frac{p_i}{V_i} O^\alpha_i \equiv - \frac{\partial}{\partial \lambda_\alpha}\ln Z = \sum_i \frac{e^{-(1/V_i)\sum_\alpha \lambda_\alpha O^\alpha_i}}{Z} O^\alpha_i.
\end{equation}

Consequently, $e^{-(1/V_i)\sum_\alpha \lambda_\alpha O_i^\alpha}/Z=p_i/V_i$ and $\varrho_\text{max-S}^{\{o_\alpha\}}=\Lambda_\text{obs}(\rho)=\sum_i p_i\Pi_i/V_i$.
Hence, the entropy production associated with the action of the channel $\Lambda_\text{obs}$ is equal to~\eqref{APPeq:obsent1}.


\section{\label{app:III}Entropy Production in System-Environment Setting}

Let us consider a system $S$ coupled to an environment $E$ prepared in a global state $\rho_{SE}$.
The reduced states of system and environment -- $\rho_S$ and $\rho_E$, respectively -- are connected with the global state $\rho_{SE}$ by the actions of the many-to-one channels $\tr_E$ and $\tr_S$: $\rho_S=\tr_E\{\rho_{SE}\}$ and $\rho_E=\tr_S\{\rho_{SE}\}$.

We assume $\rho_S$ to have been fully determined by the observer by usage of a tomographically complete set of observables $\{O_i^S\}$ with expected values $o_i^S=\tr{O_i^S\rho_S}$. 
Furthermore, we consider initially that the observer has no information whatsoever about the environment.

The trace-dual of the channel $\tr_E$ over a linear operator $O^S$ acting on the system subspace results in the extended operator $O^S\otimes\mathds{1}_E$, where $\mathds{1}_E$ is the identity over the environment subspace.
Thus, the maximum entropy state associated with the constraints $\{o_i^S\}$ reads,
\begin{equation}
 \begin{aligned}
    \varrho_\text{max-S}^{\{o_i^S\}} &= \frac{1}{Z}\exp{-\sum_i \lambda_i O_i^S\otimes\mathds{1}_E} 
    \\[0.2cm]
    &= \frac{1}{Z}\exp{-\sum_i \lambda_i O_i^S}\otimes\mathds{1}_E 
    \\[0.2cm]
    &= \frac{e^{-\sum_i \lambda_i O_i^S}}{Z_S}\otimes\frac{\mathds{1}_E}{d_E},
 \end{aligned}
\end{equation}
where
\begin{equation}
 \begin{aligned}
 Z &= \tr{\exp{-\sum_i \lambda_i O_i^S\otimes\mathds{1}_E}} 
 \\[0.2cm]
 &= \tr{\exp{-\sum_i \lambda_i O_i^S}\otimes\mathds{1}_E} 
\\[0.2cm]
&= Z_S\,d_E,   
 \end{aligned}
\end{equation}
with $Z_S=\tr{e^{-\sum_i \lambda_i O_i^S}}$ and $d_E=\tr{\mathds{1}_E}$.

Now, since the set $\{O_i^S\}$ is tomographically complete, the state $\rho_S$ is uniquely settled by the constraints $\{o_i^S\}$~\cite{Vallejos2022}.
Therefore, $\{\lambda_i\}$ must be such that $e^{-\sum_i \lambda_i O_i^S}/Z_S=\rho_S$~\cite{Vallejos2022} and $\varrho_\text{max-S}^{\{o_i^S\}}$ is given by
\begin{equation}
    \varrho_\text{max-S}^{\{o_i^S\}} = \rho_S\otimes\mathds{1}_E/d_E.
\end{equation}

Hence, the entropy produced when the observer discards $E$ -- by tracing-it-out -- is given by
\begin{equation}
    \Sigma^{\{o_i^S\}} = S(\rho_{SE}||\rho_S\otimes\mathds{1}_E/d_E).
\end{equation}

As discussed in the main text, usually in thermodynamics one considers a process in which $\rho_{SE}$ is the result of a unitary evolution from an initially decoupled state $\rho_S^0\otimes\rho_E^0$.
Moreover, in this case the observer commonly has at least partial knowledge of the initial state of the environment.
First, we assume its initial energy to be known.
Let us denote by $H_E$ and $E_E^0=\tr{\mathds{1}\otimes H_E\rho_S^0\otimes\rho_E^0}$ the Hamiltonian and initial energy of the environment.
We can look at $E_E^0$ as an additional constraint the maximum entropy state should abide to.
This leads to
\begin{equation}\label{APPeq:infostat2}
 \begin{aligned}
    \varrho_\text{max-S}^{\{o_i^S;E_E^0\}} &= \frac{1}{Z}\exp{-\sum_i \lambda_i O_i^S\otimes\mathds{1}_E-\beta_0\mathds{1}_S\otimes H_E}
    \\[0.2cm]
    &=\frac{e^{-\sum_i \lambda_i O_i^S}}{Z_S}\otimes\frac{e^{-\beta_0 H_E}}{Z_E^{\beta_0}} = \rho_S\otimes\rho_E^{\beta_0},
 \end{aligned}
\end{equation}
where $Z_E^{\beta_0}=\tr{e^{-\beta_0 H_E}}$, $\rho_E^{\beta_0}=e^{-\beta_0H_E}/Z_E^{\beta_0}$ and $\beta_0$ is the Lagrange multiplier satisfying $E_E^0=-\partial_{\beta_0}\ln Z_E^{\beta_0}$.

The entropy production associated with the observer's knowledge in this case becomes
\begin{equation}\label{APPeq:infoent1}
    \Sigma^{\{o_i^S;E_E^0\}} = S(\rho_{SE}||\rho_S\otimes\rho_E^{\beta_0}).
\end{equation}

Let us assume the initial environment state to be indeed $\rho_E^{\beta_0}$.
Then, the entropy production~\eqref{APPeq:infoent1} acquires the more thermodynamic-like form
\begin{equation}
 \begin{aligned}
    \Sigma^{\{o_i^S;E_E^0\}} &= -\tr{\rho_{SE}\ln\rho_S\otimes\rho_E^{\beta_0}} - S(\rho_{SE}) 
    \\[0.2cm]
    &= S(\rho_S) - \tr{\rho_E\ln\rho_E^{\beta_0}} - S(\rho_S^0\otimes\rho_E^{\beta_0}) 
    \\[0.2cm]
    &= \Delta S_S - \tr{(\rho_E-\rho_E^{\beta_0})\ln\rho_E^{\beta_0}} 
    \\[0.2cm]
    &= \Delta S_S +\beta_0 \tr{(\rho_E-\rho_E^{\beta_0})H_E}
    \\[0.2cm]
    &= \Delta S_S + \beta_0\Delta E_E,
 \end{aligned}
\end{equation}
where we used the invariance of the von Neumann entropy under unitary evolutions to make $S(\rho_{SE})=S(\rho_S^0\otimes\rho_E^{\beta_0})$; $\Delta S_S=S(\rho_S)-S(\rho_S^0)$ is the entropy change of the system $S$ and $\Delta E=\tr{(\rho_E-\rho_E^{\beta_0})H_E}$ the change in energy of the environment -- which is often interpreted as the heat leaving $S$.
In this scenario, Eq.~\eqref{APPeq:infoent1} was defined in~\cite{Esposito2010a,Landi2021Rev} as the entropy produced in a system put to interact with a thermal environment at inverse temperature $\beta_0$.

Crucially, Eq.~\eqref{APPeq:infoent1} can be generalized to arbitrary initial environment states.
Let $\{O_j^E\}$ denote a tomographically complete set of observables that can be used to determine the environment state $\rho_E^0$.
Let $o_j^{E_0}=\tr{O_j^E\rho_E^0}$ be the expected value of $O_j^E$ in this state. 
The maximum entropy state abiding to the constraints $\{o_i^S;o_j^{E_0}\}$ is given by
\begin{equation}\label{eq:maxSQSTOpen}
 \begin{aligned}
    \varrho_\text{max-S}^{\{o_i^S;o_j^{E_0}\}} &= \frac{1}{Z}\exp{-\sum_i \lambda_i O_i^S\otimes\mathds{1}_E-\sum_j \xi_j^0 \mathds{1}_S\otimes O_j^E} 
    \\[0.2cm]
    &= \frac{e^{-\sum_i \lambda_i O_i^S}}{Z_S}\otimes\frac{e^{-\sum_j \xi_j^0 O_j^E}}{Z_E^0} = \rho_S\otimes\rho_E^0,
 \end{aligned}
\end{equation}
where $Z_E^0=\tr{e^{-\sum_j \xi_j^0 O_j^E}}$ and we used that the Lagrange multiplies $\{\xi_j^0\}$ must be such that $e^{-\sum_j \xi_j^0 O_j^E}/Z_E^0=\rho_E^0$.

The entropy production thus becomes
\begin{equation}\label{APPeq:infoent2}
    \Sigma^{\{o_i^S;o_j^{E_0}\}} = S(\rho_{SE}||\rho_S\otimes\rho_E^0) = I(\rho_{SE}) + S(\rho_E||\rho_E^0),
\end{equation}
where $I(\rho_{SE})=S(\rho_{SE}||\rho_S\otimes\rho_E)=S(\rho_S) + S(\rho_E) - S(\rho_{SE})$ is the quantum mutual information in $\rho_{SE}$.
Equation~\eqref{APPeq:infoent2} is defined in~\cite{Esposito2010a,Landi2021Rev} as the entropy produced when a system $S$ is put to interact with an environment initially in a state $\rho_E^0$ and the latter is discarded at the end of the interaction.
The last equality in~\eqref{APPeq:infoent2} shows this entropy production is related to the observer's lack of information about the correlations between $S$ and $E$ in the final state $\rho_{SE}$ -- computed by $I(\rho_{SE})$ -- and absence of information about the final environment state $\rho_E$.

Next, let us also consider the situation where the environment is not discarded: at the end of the interaction the observer can perform local measurements on both $S$ and $E$.
Let $\{o_i^S=\tr{O_i^S\rho_S}\}$ and $\{o_j^E=\tr{O_j^E\rho_E}\}$ be the sets of constraints that completely specify the final (reduced) states of system $\rho_S$ and environment $\rho_E$.
The maximum entropy state associated with these constraints is given by
\begin{equation}
 \begin{aligned}
    \varrho_\text{max-S}^{\{o_i^S;o_j^{E}\}} &= \frac{1}{Z}\exp{-\sum_i \lambda_i O_i^S\otimes\mathds{1}_E-\sum_j \xi_j \mathds{1}_S\otimes O_j^E} 
    \\[0.2cm]
    &= \frac{e^{-\sum_i \lambda_i O_i^S}}{Z_S}\otimes\frac{e^{-\sum_j \xi_j O_j^E}}{Z_E} = \rho_S\otimes\rho_E,
 \end{aligned}
\end{equation}
where $Z_E=\tr{e^{-\sum_j \xi_j O_j^E}}$ and we used that $\{\xi_i\}$ must be such that $e^{-\sum_j \xi_j O_j^E}/Z_E = \rho_E$.

The entropy production in this case amounts to
\begin{equation}\label{APPeq:entprod5}
    \Sigma^{\{o_i^S;o_j^E\}} = S(\rho_{SE}||\rho_S\otimes\rho_E) = I(\rho_{SE}),
\end{equation}
and is solely related with the lack of knowledge of the observer about the correlations in the global state $\rho_{SE}$.


\section{\label{app:IV}Entropy Production with Coarse-Grained Measurement on the Environment}

Let us continue to consider the thermodynamic scenario of a system and environment initially in an uncorrelated state $\rho_S^0\otimes\rho_E^0$ evolving unitarily to $\rho_{SE}$ at time $t$.

Let $H_E = \sum_{j\mu}\epsilon_{j\mu}^E |\epsilon_{j\mu}^E\rangle\langle\epsilon_{j\mu}^E|$ denote the (time-independent) environment Hamiltonian and $\{|s_i^t\rangle\}$ denote the orthonormal eigenbasis of the system local state $\rho_S(t)=\tr_E\{\rho_{SE}(t)\}=\sum_i s_i^t|s_i^t\rangle\langle s_i^t|$ at time $t$.
Moreover, let $\{\Pi_j^E=\sum_{\mu} |\epsilon_{j\mu}^E\rangle\langle\epsilon_{j\mu}^E| \}$ be a complete, $\sum_j \Pi_j^E=\mathds{1}_E$, and orthogonal, $\Pi_j^E\Pi_{j'}^E=\delta_{jj'}\Pi_j^E$, set of energy-projectors with ranks $V_j^E=\tr{\Pi_j^E}$.
We assume an observer who can, at any time $t$, perform the joint measurement characterized by the operator set $\{|s_i^t\rangle\langle s_i^t|\otimes\Pi_j^E\}$.
Hence, the measurement is coarse on the energy of the environment.
Finally, we assume the initial system-environment state to be of the form $\rho_{SE}^0=\rho_S^0\otimes\rho_E^0=\sum_i s_i^0 |s_i^0\rangle\langle s_i^0|\otimes\sum_j p_j^0\Pi_j^E/V_j^E$ such that
\begin{equation}
    S_\text{obs}^{\{|s_i^0\rangle\langle s_i^0|\otimes\Pi_j^E\}}(\rho_{SE}^0) = S(\rho_{SE}^0).
\end{equation}

The measurement $\{|s_i^t\rangle\langle s_i^t|\otimes\Pi_j^E\}$ at time $t$ allows the determination of the probabilities $p_{ij}=\tr{|s_i^t\rangle\langle s_i^t|\otimes\Pi_j^E\rho_{SE}(t)}$ of finding the system in the eigenstate $|s_i^t\rangle$ while the environment is in the energy shell defined by the projector $\Pi_j^E$.
The maximum entropy state compatible with the knowledge of these probabilities is given by
\begin{equation}
 \begin{aligned}
    \varrho_\text{max-S}^{\{p_{ij}\}} &= \frac{1}{Z}\exp{-\sum_{ij}\xi_{ij}|s_i^t\rangle\langle s_i^t|\otimes\Pi_j^E} 
    \\[0.2cm]
    &= \sum_{ij} \frac{e^{-\xi_{ij}}}{Z}|s_i^t\rangle\langle s_i^t|\otimes\Pi_j^E,
 \end{aligned}
\end{equation}
where
\begin{equation}
    Z=\tr{\exp{-\sum_{ij}\xi_{ij}|s_i^t\rangle\langle s_i^t|\otimes\Pi_j^E}}=\sum_{ij} V_j^Ee^{-\xi_{ij}},
\end{equation}
and the Lagrange multipliers $\{\xi_{ij}\}$ must be such that
\begin{equation}
    p_{ij} \equiv -\frac{\partial}{\partial\xi_{ij}}\ln Z = V_j^E\frac{e^{-\xi_{ij}}}{Z}.
\end{equation}
Thus, $e^{-\xi_{ij}}/Z=p_{ij}/V_j^E$ and
\begin{equation}\label{APPeq:ObsStat2}
    \varrho_\text{max-S}^{\{p_{ij}\}} = \sum_{ij} p_{ij}|s_i^t\rangle\langle s_i^t|\otimes\Pi_j^E/V_j^E.
\end{equation}

The entropy production associated with this observer's knowledge is thus
\begin{equation}\label{APPeq:obsent2}
\begin{aligned}
    \Sigma^{\{p_{ij}\}} &= S(\rho_{SE}||\varrho_\text{max-S}^{\{p_{ij}\}})
    \\[0.2cm]    
    &= S_\text{obs}^{\{|s_i^t\rangle\langle s_i^t|\otimes\Pi_j^E\}}(\rho_{SE}) - S_\text{obs}^{\{|s_i^0\rangle\langle s_i^0|\otimes\Pi_j^E\}}(\rho_{SE}^0),
\end{aligned}
\end{equation}
where we used that $S(\rho_{SE})=S(\rho_{SE}^0)=S_\text{obs}^{\{|s_i^0\rangle\langle s_i^0|\otimes\Pi_j^E\}}(\rho_{SE}^0)$ and
\begin{equation}
    S_\text{obs}^{\{|s_i^t\rangle\langle s_i^t|\otimes\Pi_j^E\}}(\rho_{SE}) = -\sum_{ij} p_{ij}\ln (p_{ij}/V_j^E)
\end{equation}
is the observational entropy of $\rho_{SE}$.

Let us also consider the case of an observer performing \emph{local} measurements on the system and environment characterized by the operator sets $\{|s_i^t\rangle\langle s_i^t|\otimes\mathds{1}_E\}$ and $\{\mathds{1}_S\otimes\Pi_j^E\}$.
In this case, the two measurements allow the observer to determine the (uncorrelated) probabilities $s_i^t=\tr\{|s_i^t\rangle\langle s_i^t|\otimes\mathds{1}_E\rho_{SE}\}=\tr_S\{|s_i^t\rangle\langle s_i^t|\rho_S\}$ of finding the system in eigenstate $|s_i^t\rangle$ and $p_j^E=\tr\{\mathds{1}_S\otimes\Pi_j^E\rho_{SE}\}=\tr_E\{\Pi_j^E\rho_E\}$ -- where $\rho_E=\tr_S\{\rho_{SE}\}$ is the local environment state at time $t$ -- of finding the environment in the energy shell associated with $\Pi_j^E$.
The maximum entropy state constrained by $\{s_i^t;p_j^E\}$ is given by
\begin{equation}
 \begin{aligned}
    \varrho_\text{max-S}^{\{s_i^t;p_j^E\}} &= \frac{1}{Z}\exp{-\sum_i \xi_i^S |s_i^t\rangle\langle s_i^t|\otimes\mathds{1}_E - \sum_j \xi_j^E \mathds{1}_S\otimes\Pi_j^E } 
    \\[0.2cm]
    &= \sum_i\frac{e^{-\xi_i^S}}{Z_S'}|s_i^t\rangle\langle s_i^t|\otimes\sum_j\frac{e^{-\xi_j^E}}{Z_E'}\Pi_j^E,
 \end{aligned}
\end{equation}
where $Z_S'=\tr_S\{e^{-\sum_i \xi_i^S |s_i^t\rangle\langle s_i^t| } \} = \sum_i e^{-\xi_i^S}$ and $Z_E'= \tr_E\{e^{-\sum_j \xi_j \Pi_j^E}\} = \sum_j V_j^Ee^{-\xi_j^E}$.
The Lagrange multipliers $\{\xi_i^S\}$ and $\{\xi_j^E\}$ are such that
\begin{IEEEeqnarray}{rCl}
    s_i^t &=& - \frac{\partial}{\partial\xi_i^S}\ln Z_S' = \frac{e^{-\xi_i^S}}{Z_S'},
    \\[0.2cm]
    p_j^E &=& - \frac{\partial}{\partial\xi_j^E}\ln Z_E' = \frac{e^{-\xi_j^E}}{Z_E'}V_j^E.
\end{IEEEeqnarray}
This means $e^{-\xi_i^S}/Z_S'=s_i^t$, $e^{-\xi_j^E}/Z_E'=p_j^E/V_j^E$ and
\begin{equation}
 \begin{aligned}
    \varrho_\text{max-S}^{\{s_i^t;p_j^E\}} &= \sum_i s_i^t|s_i^t\rangle\langle s_i^t|\otimes\sum_j p_j^E\Pi_j^E/V_j^E 
    \\[0.2cm]
    &= \rho_S\otimes \sum_j p_j^E\Pi_j^E/V_j^E. 
 \end{aligned}
\end{equation}

The entropy production in this case becomes
\begin{equation}\label{APPeq:obsent3}
    \Sigma^{\{s_i^t;p_j^E\}} = S(\rho_{SE}||\varrho_\text{max-S}^{\{s_i^t;p_j^E\}}) = \Delta S_\text{obs}^S + \Delta S_\text{obs}^E,
\end{equation}
where
\begin{IEEEeqnarray}{rCl}
    \Delta S_\text{obs}^S &=& S_\text{obs}^{\{|s_i^t\rangle\langle s_i^t|\}}(\rho_S) - S_\text{obs}^{\{|s_i^0\rangle\langle s_i^0|\}}(\rho_S^0) 
    \nonumber\\[0.1cm]
    &=& S(\rho_S) - S(\rho_S^0),\IEEEeqnarraynumspace
    \\[0.5cm]
    \Delta S_\text{obs}^E &=& S_\text{obs}^{\{\Pi_j^E\}}(\rho_E) - S_\text{obs}^{\{\Pi_j^E\}}(\rho_E^0)
\end{IEEEeqnarray}
are the changes in observational entropy of the system and environment, respectively, and we used that
\begin{equation*}
 \begin{aligned}
    S(\rho_{SE}) &= S(\rho_{SE}^0) = S_\text{obs}^{\{|s_i^0\rangle\langle s_i^0|\otimes\Pi_j^E\}}(\rho_{SE}^0) 
    \\[0.2cm]
    &= S_\text{obs}^{\{|s_i^0\rangle\langle s_i^0|\}}(\rho_{S}^0) + S_\text{obs}^{\{\Pi_j^E\}}(\rho_{E}^0).
 \end{aligned}
\end{equation*}

The difference between the two entropy productions~\eqref{APPeq:obsent3} and~\eqref{APPeq:obsent2} is given by~\cite{Strasberg2021}
\begin{equation}\label{APPeq:obsentdif}
    \Sigma^{\{s_i^t;p_j^E\}} - \Sigma^{\{p_{ij}\}} = I_c(\{p_{ij}\}) \geqslant 0,
\end{equation}
where
\begin{equation}
    I_c(\{p_{ij}\}) = \sum_{ij} p_{ij} \ln \frac{p_{ij}}{s_i^tp_j^E} 
\end{equation}
is the classical mutual information.
Equation~\eqref{APPeq:obsentdif} shows how knowledge of the correlations between the eigenstates $|s_i^t\rangle$ of the system and the energy shells $\Pi_j^E$ of the environment reduce the total amount of entropy production.


\section{\label{app:V}Entropy Change in One-to-One Maps}
As discussed in the main text, for channels $\Lambda$ that establish a one-to-one relation between its input state $\rho$ and output $\Lambda(\rho)$, we have $\varrho_\text{max-S}=\rho$.
According to our definition, $\Sigma = S(\rho||\varrho_\text{max-S})$, this leads to a zero entropy production $\Sigma=0$.
Nonetheless, we can still use this definition to predict the change in von Neumann entropy of the system in these scenarios as follows.

We can think of these channels as resulting from the interaction of the system $S$ with an environment $A$.
Let $\rho_S\otimes|0\rangle_A\langle0|$ be the initial joint state of $S$ and $A$ and $U_\Lambda$ be the unitary generating the channel $\Lambda$, such that
\begin{equation}
    \Lambda(\rho_S)=\tr_A\{U_\Lambda(\rho_S\otimes|0\rangle_A\langle0|)U_\Lambda^\dagger\}.
\end{equation}
As seen in Eq.~\eqref{APPeq:infoent2}, the tracing-out of $A$, assuming its initial \emph{pure state} to be \emph{known}, leads to an entropy production given by
\begin{equation}
\begin{aligned}
    \Sigma &= S(U_\Lambda(\rho_S\otimes|0\rangle_A\langle0|)U_\Lambda^\dagger||\Lambda(\rho_S)\otimes|0\rangle_A\langle0|) 
    \\[0.2cm]
    &= S(\Lambda(\rho_S)) - S(\rho_S),
\end{aligned}
\end{equation}
which is the change in von Neumann entropy of the system after passing through the channel $\Lambda$.

\bibliography{library1,library2}

\end{document}